\documentclass{aa}
\usepackage{txfonts}
\usepackage{natbib}
\usepackage{graphicx}
\usepackage{url}
\bibpunct{(}{)}{;}{a}{}{,}

\begin{document}

\title{Radiative transfer on hierarchial grids}
\author{T. Lunttila\and M. Juvela}
\institute{Department of Physics, P.O. Box 64, FI-00014, University of Helsinki, Finland\\
\email{tuomas.lunttila@helsinki.fi}}

\date{Received / Accepted}

\abstract {Continuum radiative-transfer simulations are necessary for the interpretation of observations of dusty astrophysical objects and for relating the results of magnetohydrodynamical simulations to observations. The calculations are computationally difficult, and simulations of objects with high optical depths in particular require considerable computational resources.}
{Our aim is to show how radiative transfer calculations on adaptive three-dimensional grids can be accelerated.}
{We show how the hierarchial tree structure of the model can be used in the calculations. We develop a new method for calculating the scattered flux that employs the grid structure to speed up the computation. We describe a novel subiteration algorithm that can be used to accelerate calculations with strong dust temperature self-coupling. We compute two test models, a molecular cloud and a circumstellar disc, and compare the accuracy and speed of the new algorithms against existing methods.}
{An adaptive model of the molecular cloud with fewer than 8 \% of the cells in the uniform grid produces results in good agreement with the full resolution model.  The relative root-mean-square (RMS) error of the surface brightness is $\la 4$ \% at all wavelengths, and in regions of high column density the relative RMS error is only $\sim 10^{-4}$. Computation with the adaptive model is faster by a factor of $\sim 5$. Our new method for calculating the scattered flux is faster by a factor of about four in large models with a deep hierarchy structure, when images of the scattered light are computed towards several observing directions. The efficiency of the subiteration algorithm is highly dependent on the details of the model. In the circumstellar disc test the speed-up is a factor of two, but much larger gains are possible. The algorithm is expected to be most beneficial in models where a large number of small, dense regions are embedded in an environment of low mean density.}
{}
\titlerunning{Radiative transfer on hierarchial grids}
\maketitle

\section{Introduction}
Radiative transfer modelling is an indispensable tool in the interpretation of observations of dusty astrophysical objects such as circumstellar discs \citep[e.g.,][]{Acreman2010}, molecular cloud cores \citep[e.g.,][]{Steinacker2005}, spiral galaxies \citep[e.g.,][]{deLooze2012}, and galaxy mergers \citep[e.g.,][]{Hayward2011}. Solving the radiative transfer equation is numerically a very difficult problem, requiring considerable computational resources. It is sometimes possible to use one-dimensional or two-dimensional geometries, but for a realistic representation of inhomogenous structures such as turbulent molecular clouds, a fully three-dimensional (3D) model is needed. Moreover, an accurate description of the structure often requires the inclusion of a large variety of scales. For instance, a model of a circumstellar disc may require a resolution of $\sim R_{\sun}$ near the star, while, to include the whole disc, the total extent of the model needs to be several hundred AU. On a uniform cartesian grid, such a model would comprise hundreds of billions cells. Furthermore, the radiative transfer problem needs to be solved at several wavelengths and iteratively, if the dust is hot and the model is not optically thin, to calculate the thermal dust emission self-consistently.

To reduce computational cost, several 3D radiative-transfer codes have been developed with support for adaptive resolution, i.e., the possibility of using a higher resolution in some parts of the model. With adaptive resolution, it is possible to use the finest resolution only where necessary, thereby reducing the number of cells in the model in some cases by many orders of magnitude. With cartesian grids, the most commonly used structure has been the oct-tree \citep[e.g.,][]{Jonsson2006, Acreman2010}. In an oct-tree, every model cell can be divided into eight subcells, which can then be divided further. A completely different approach was chosen by \citet{Ritzerveld2006}, who dispenses with the cartesian grid and moved the photons along the edges of Delaunay triangles in a point cloud.

Regardless of the method used to calculate the radiation field, iteration is needed in cases where the dust self-coupling is significant. The simplest and most commonly used method is the $\Lambda$ iteration, but this suffers from very slow convergence in models with a high optical depth. Convergence can be improved with accelerated $\Lambda$ iteration (ALI) at the cost of increased computer memory requirements and the additional computation required at each iteration step \citep{Cannon1973,Rybicki1991}. Accelerated lambda iteration was reformulated for use with the Monte Carlo methods in \citet{Hogerheijde2000}, where it was called an accelerated Monte Carlo method \citep[see also][]{Juvela1999}. These methods are based on treating separately the part of the radiation emitted by a cell that is absorbed in the same cell or, in some variations of the method, in its immediate neighbourhood. Nevertheless, models with optical depths of several thousand, such as dense circumstellar envelopes, can require tens of iterations even when using ALI \citep{Juvela2005}. For a large model, the computation of a single iteration can be very time-consuming, making solving the full problem infeasible.

The programme described in this article is based on the Monte Carlo method. The main difference from other Monte Carlo radiative-transfer codes is the use of a hierarchial tree structure of nested grids, closely resembling that employed in patch-based adaptive mesh refinement (AMR) hydrodynamics codes. Although hierarchial grids have been used in radiative transfer calculations before ~\citep[e.g.,][]{Robitaille2011}, the method described here differs from the previous ones in some key aspects. In the Monte Carlo simulation, the programme works grid-by-grid, moving to the next only after all photon packages in a grid have been processed instead of following one photon package at a time through the whole model. The most important new feature is the possibility of using subiterations, i.e., iterating separately those parts of the model that suffer from slow convergence. Although the current implementation uses the Monte Carlo method to compute the formal solution, the subiteration algorithm is independent of the solution method. The programme described here has already been used in the study of molecular cloud cores \citep{Malinen2011, Juvela2011} and galaxy mergers (Karl et al., in preparation).

We describe how the Monte Carlo radiation transfer is performed on a hierarchial grid in Sect. 2, while the use of subiterations is explained in Sect. 3. In Sect. 4, we present results from some tests of the new method and compare them with a radiative transfer code that uses a regular 3D grid. Section 5 discusses possible future extensions, and Section 6 presents the conclusions.

\section{Formal solution}
Because of dust self-heating and the long-distance radiative couplings between regions, the continuum radiative-transfer problem is both non-linear and non-local. The full problem is usually solved using iterative methods, although some 'immediate re-emission' codes \citep[see][]{Bjorkman2001} avoid the explicit iteration process. Each step of the iteration typically requires the solution of a linear radiative-transfer problem (i.e., calculating the formal solution) with known dust thermal emission from the previous iteration.

Our current implementation employs the Monte Carlo method for solving the linear transfer problem. The basic use of the Monte Carlo method in radiative transfer calculations, i.e., tracking photon packages, is well-established (see, e.g., \citet{Whitney2011} and references therein) and is not described here in detail. The part of the code that performs the Monte Carlo sampling in each grid is identical to the uniform-grid radiative-transfer programme described in \citet{Juvela2003} and \citet{Juvela2005}. Therefore, the following discussion is limited to the parts of the programme involving the interaction between different grids.

\subsection{Structure of the model}
The dust density distribution is discretised on an adaptive mesh of nested grids. The grids form a tree structure, where each individual grid is a node in a tree. Figure 1 shows an example of the hierarchy structure of a simple model. There is a single root grid (grid 0) that contains the whole simulation volume. Some parts of the root grid can belong to a subgrid that has a finer resolution; these subgrids are children of the root grid and the root grid is their parent. The subgrids can have their own children (i.e., subgrids), which can have their own children, continuing until there is no need for further refinement. The depth of a grid is the length of the path joining the root grid to the grid; the depth of the root node is 0. A level of the hierarchy consists of all grids at the same depth. The vertices of a subgrid are always restricted to integer co-ordinates in the parent grid, so that a subgrid always fully replaces an integer number of cells of its parent grid. Two children of a parent grid cannot overlap. In the subgrid, the linear size of a cell is a reciprocal of an integer (usually one half) times the size of a parent grid cell.

\begin{figure}
  \resizebox{\hsize}{!}{\includegraphics{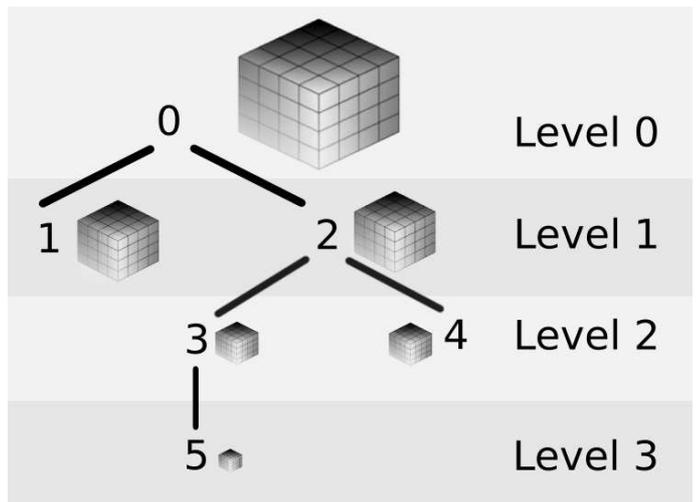}}
  \caption{Hierarchy structure of a simple model.}
\end{figure}

The structure of the grid hierarchy is described in a hierarchy file that lists the location and size of every grid. In addition to the hierarchy file, there are one or more files for each grid that describe the density structure (possibly for several dust populations) and, for example in simulations of dusty galaxies, the stellar emission. Basic simulation parameters such as the size of the model in physical units and the description of the required output maps and spectra are set in an initialisation file. Because the programme was initially used mainly for running radiative transfer simulations in snapshots produced by the AMR (magneto)hydrodynamics code ENZO \citep{OShea2004, Collins2010}, we chose to use its format for the hierarchy file. We also wrote tools to convert data given on a uniform grid or as a smoothed particle hydrodynamics snapshot into the format used by the programme.

Before the simulation starts, a subgrid indentification array is constructed for every grid of the model. The array shows for each cell in the grid whether the cell belongs to a subgrid and, if it does, the number of that subgrid. It is used during the simulation to quickly check whether a photon package has encountered a subgrid. Because this needs to be done every time that the package crosses a cell, it is important that it can be done with a fast table lookup instead of a time-consuming search among the grids listed in the hierarchy file.

\subsection{The grid boundaries}
The external boundary of every grid has two arrays for storing photon packages: one for the packages that enter the grid from the outside (photons going inwards, \textit{inwards} array) and one for the packages exiting the grid through its outer boundary (photons going outwards, \textit{outwards} array). If a package exits the grid via the outer boundary, it is stored in the grid's \textit{outwards} array. If a package enters a subgrid instead, it is stored in that subgrid's \textit{inwards} array.

When a package reaches a grid boundary, the programme stores the number of photons in the package, the direction cosines, and the position where the package crosses the boundary. Some additional numbers are saved at the boundary when subiterations are used (see Sect 3.4). Because the number of photon packages that need to be saved on a grid boundary cannot be known when the simulations starts, the size of the arrays must be changed dynamically during the simulation.

\subsection{Solving the linear problem}
The simulation begins with the creation of photon packages that describe the radiation from sources inside the model volume, such as stars or dust thermal emission. The packages can be polychromatic so that a single Monte Carlo ray represents several wavelengths \citep{Jonsson2006, Jonsson2010}. A package is followed step by step through the grid until it exits the grid in which it was created either by encountering its outer boundary or by entering one of its subgrids. The packet is stored in the appropriate boundary array -- the grid's \textit{outwards} array if the packet reaches the outer edge of the grid, and a child grid's \textit{inwards} array if the packet encounters a subgrid. The simulation proceeds with the next package emitted inside the grid. After all packages are processed, the programme switches to the next grid and the same process is repeated. The tracking of a photon package inside a grid is done using local grid coordinates, where the linear size of each cell is one unit. This allows us to use exactly the same routine for following the packages at all levels without the need to adjust for different cell sizes. The global coordinates that refer to the position in the root grid's coordinate system are only used when a package crosses a boundary.

When the internal emission from all grids has been processed, the next step is to transport the packages stored on the grid boundaries. This is started from the root grid, whose \textit{inwards} array initially contains the photon packages entering the simulation volume from the outside. These represent the background radiation, e.g., the interstellar radiation field. The packages are followed until they exit the grid either through the outer boundary, or by entering a subgrid, and they are stored in the corresponding boundary array. The package is removed from the \textit{inwards} array, and the simulation proceeds with the next package. After all packages are processed and the \textit{inwards} array is empty, the simulation continues with the next grid. The simulation moves inwards so that a child grid is always processed after its parent. Therefore, a child grid's \textit{inwards} array can contain packages both from sources within the parent grid and from external radiation reaching the parent's \textit{inwards} array. After all grids have been processed, there are no packages in any \textit{inwards} array. The simulation continues with \textit{outwards} arrays, starting from the deepest hierarchy level and moving outwards. Packages on the root grid's \textit{outwards} boundary escape the simulation volume and are deleted. Because new packages are started only at the beginning of the simulation when the radiation sources are processed, the total number of packages in the left simulation decreases during the computation.\footnote{This is not the case if package splitting at the grid boundaries is used.} The computation continues by alternatingly processing the hierarchy inwards and outwards, until there are no packages left. A detailed example of the process is given in Appendix A.

We use the continuous absorption method described in \citet{Lucy1999}, so that a package is normally removed from the simulation only if it exits through the outer boundary of the root grid. However, for computational efficiency it is sometimes advantageous to terminate packages when they have lost most of their photons. This is done by using a Russian roulette scheme. If the number of photons in a package falls below the limit $n$, it is deleted with probability $p$. If the packet survives, the number of photons in the package is multiplied by $(1-p)^{-1}$. The parameters $n$ and $p$ controlling the Russian roulette can be set in the initialisation file.

The programme also includes the possibility of using package splitting to improve sampling in subgrids. Several photon packages are started for every package stored in the \textit{inwards} array, with the number of photons in each divided correspondingly. When the package splitting is used, it is usually also necessary to employ a Russian roulette scheme at the \textit{outwards} boundary to limit the number of packages that later continue to the parent grid, so that if a package entering the grid is divided into $N$ parts, then the packages exiting the grid have only a $1/N$ probability of surviving.

\subsection{Peel-off}
The peel-off method \citep{Yusef-Zadeh1984} is a technique for calculating images of the scattered light that produces higher signal-to-noise ratio images compared to the na\"{\i}ve Monte Carlo simulation. In the peel-off method, one calculates after each scattering the fraction of photons that are scattered towards one or more observers and escape the model. The escaped photons are used to construct the images of scattered light. The use of the peel-off method increases the computational cost of each scattering event, but because every scattering contributes to the final image, the number of photon packages needed for a high quality image is reduced significantly. The calculation of the scattered light maps is done as the final post-processing step after the dust temperature distribution has been solved with the radiative-transfer computation.

We use a novel variation of the technique that accelerates the computation. Before the start of the simulation, the outer boundaries of selected subgrids are divided into small tiles. The optical thickness is calculated from the centre of each tile to the outer border of the model for each observer direction, and the optical thickness values are stored into a table. During the peel-off calculation, it is necessary to follow the package only until it meets the boundary of a subgrid for which the extinction table has been calculated. The total extinction is calculated by adding the extinction to the subgrid boundary to the precalculated value that was saved in the table. The use of precalculated extinction tables can reduce the computation time per photon package to a fraction of its original length at the cost of the need to perform more preliminary calculation at the start of the simulation.

\section{Local iterations}
\subsection{The basic algorithm}
The radiative transfer equation can be formally written as
\begin{equation}
\label{rtequt}
\begin{array}{ccc}
J& = &\Lambda S + J_0\\
S& = &f(J),
\end{array}
\end{equation}
where $J$ is the radiation field, $S$ is the source function, and $J_0$ is the radiation field due to constant sources such as stars. Operator $\Lambda$ is a linear mapping from the source function to the radiation field and $f$ is the function that relates the radiation absorbed by the dust to its thermal emission. If we assume, as is usually done in dust radiative-transfer simulations, that the opacity is independent of the dust temperature, $\Lambda$ does not depend on $J$. Because the system of equations \ref{rtequt} has $N_{\mathrm{cells}} N_{\mathrm{freq}}$ unknowns, solving it with, e.g., Newton-Raphson iteration is not possible for large models. In particular, the matrix representing the discretised $\Lambda$ operator has $N_{\mathrm{freq}}N_{\mathrm{cells}}^2$ (possibly) non-zero elements, needing hundreds of terabytes of storage for a model with several million cells.

The system of equations in Eq. \ref{rtequt} can be solved without explicitly constructing the $\Lambda$ operator by using the $\Lambda$ iteration
\begin{equation}
\label{liequt}
\begin{array}{ccc}
J_{n+1} &= &\Lambda S_n + J_0\\
S_{n+1} &= &f(J_{n+1}).
\end{array}
\end{equation}
Each iteration step entails both solving a linear radiative-transfer problem for a given source function using, e.g., the Monte Carlo method, and calculating the source function from the radiation field using the dust model. In large models and with sophisticated dust models, both of these calculations are computationally expensive, limiting the ability to solve the radiative transfer equation in slowly converging models.

In a model where individual cells are optically thick, most of the radiation emitted in a cell is absorbed in the same cell, and therefore does not contribute to the net energy transfer between different cells. This means that in the matrix representing the $\Lambda$ operator, the entries on the main diagonal of the matrix are large compared to the other entries, leading to a very slow convergence of the basic $\Lambda$ iteration \citep{Rybicki1991}. In the accelerated $\Lambda$ iteration, the problem of slow convergence is mitigated by explicitly treating the diagonal part of the $\Lambda$ operator. The operator is written as $\Lambda=\Lambda_0+\Lambda_1$, where $\Lambda_0$ is a diagonal matrix. The iteration is then run as
\begin{equation}
\label{aliequt}
\begin{array}{ccc}
J_{n+1} &= &\Lambda_0 S_{n+1}+\Lambda_1 S_n + J_0\\
S_{n+1} &= &f(J_{n+1}).
\end{array}
\end{equation}
Every step of the iteration involves the solution of a non-linear system of equations. However, because $\Lambda_0$ is diagonal, the full system decouples into $N_{\mathrm{cells}}$ separate systems, each with $N_{\mathrm{freq}}$ unknowns. Furthermore, because only the diagonal part of $\Lambda$ is needed, storage requirements are much lower. Instead of a diagonal $\Lambda_0$, it is possible to use more complex operators that better approximate $\Lambda$. This accelerates convergence, but requires more storage and computation for each iteration step \citep{Juvela2005}.

The optically thick, slowly converging regions may comprise only a small part of the simulation volume. The grid structure allows us to exploit this fact. Instead of solving the system of equations in Eqs. \ref{liequt} or \ref{aliequt} in the whole simulation volume for each iteration, we use subiterations, i.e. take an iteration step only in the slowly converging grids. If the subiterations are done only in a small part of the model, they are faster than full iterations (see Appendix B for a detailed discussion). If the bulk of the model is optically thin, only a small number of iterations of the full model are needed in addition to the subiterations of the densest grids.

\subsection{Implementation}
The simulation begins with the computation of the radiation field produced by constant sources such as stars and the external radiation. The radiation field due to constant sources is saved for each cell in the model. Because we assume that the dust opacity does not depend on temperature, this calculation does not need to be repeated and only the dust thermal emission needs to be recalculated in the subsequent iterations. In the following iterations, the dust emission for all cells in the grids that are included in the subiteration is computed using the previously calculated radiation field. Thereafter, the Monte Carlo radiative-transfer simulation is run to calculate the radiation field due to dust emission.

The main difficulty in the implementation of the subiteration algorithm is ensuring that the total radiation field in a cell is always calculated using the data from the most recent subiterations. Because the grids may have had a different number of iterations, this requires careful tracking of the radiation from different grids. To enable this without requiring a very large amount of computer memory, we do not permit the inclusion of an arbitrary set of grids in a subiteration. A subiteration must instead be done in a complete subtree. In particular, this means that taking a subiteration step with a single grid is allowed only if the grid does not have any children. We present a detailed description of the algorithm in Appendix B.

We chose not to include the calculation of dust emission (i.e., evaluating $f(J)$) in the radiative-transfer programme. The total radiation field in each cell is written to a computer disc and a separate programme is called to calculate the new source function. Therefore, any dust modelling code can be easily used with our radiative transfer program. If a dust model with several dust populations, stochastically heated non-local thermodynamic equilibrium (LTE) particles, and polyaromatic hydrocarbons (PAH) emission is used, the time spent calculating the dust emission can be much longer than that spent in the radiation transfer step. In these cases, it is often necessary to use acceleration methods \citep[e.g.,][]{Juvela2003,Baes2011} to calculate the dust emission. There is some overhead owing to the storage of data on the disc instead of computer memory. However, in all our tests the time spent in disc input/output (I/O) is less than 4 \% of the total running time. Moreover, in simulations with a large model and a dense frequency grid the radiation field data can take more than a hundred gigabytes and it may be impossible to store the data in the memory.

We also included the possibility of using ALI with a diagonal approximate $\Lambda$ operator. Whether ALI is used can be chosen separately for each grid. For instance, one can choose to use ALI in only a few grids with the highest optical thickness. The use of ALI accelerates the convergence in optically thick grids, thereby reducing the number of iterations and saving computer time. Furthermore, by accelerating the convergence ALI makes it easier to determine whether a grid has converged. For the details of implementing ALI, we refer to \citet{Juvela2005}.

\subsection{Automatic iteration}
The order in which different subtrees are processed can be defined by the user. However, this is impractical in models with a large number of grids, and it is better to let the programme automatically determine which parts of the model need more subiterations. This is done by tracking the energy balance, i.e., the difference between energy absorbed and emitted by the dust, at the level of individual cells as well as entire grids, and by choosing the grids with the largest imbalance for the next iteration. The iteration process is terminated when the energy balance has been attained with the required accuracy.

Monitoring the convergence in a single grid can be done by computing the difference in energy absorbed and emitted by the dust grains in each cell. Even if at a given point during the iteration, there is a perfect balance between the absorbed and emitted energy in every cell of the model, the iteration process as a whole has not necessarily converged. For instance, if the only radiation source in the model is within a very optically thick grid, energy balance for every cell in the model could be reached by iterating only the grid containing the source. However, subiterations of larger subtrees are necessary to balance the transfer of energy between different grids. The programme tracks the energy flow from each grid to its parent. If the energy flux into a grid from its children has changed significantly (e.g., the change is an appreciable fraction of the total energy absorbed in the grid) since the last iteration where the grid was included, the grid is tagged for the next iteration.

The user can adjust the parameters controlling the iteration process, for instance by choosing whether the criterion determining the next subtrees is the maximum imbalance between the absorbed and emitted energy by a cell, or a weighted average over all cells in the grid. It is usually inefficient to include in the next iteration only the subtree starting from the grid with the largest energy imbalance. For instance, if a grid has only slightly smaller energy imbalance than one of its children, it is better to also include that grid in the next iteration. Otherwise, it would very likely be incorporated in a later iteration, and the earlier iteration of its subtree could be at least partly redundant. The user can set the parameter controlling how close to the largest found energy imbalance the grids need to be in order to be included in the next iteration.

The statistical noise inherent to the Monte Carlo method causes fluctuations even if the iteration has converged. In some cases, especially if the number of photon packages in the simulation is low, this can cause the iteration to get stuck in a converged subtree, while other parts of the model may not have yet converged. This may be alleviated by following the convergence. If there is no improvement, the iteration may have reached the noise limit, and the algorithm forces another choice of a rootgrid for the next iteration. However, it is important not to terminate the iteration prematurely, because for optically thick systems the convergence can be slow.

\subsection{Computational issues}
Although the grid-by-grid processing seems complicated compared to simply following one photon package at a time through the whole model, it introduces little computational overhead. In a model with $\approx 4\times 10^6$ cells and $\approx 300$ grids, i.e., a relatively large number of small grids, only $\approx 3$ \% of the CPU time is spent in the transfer of photon packages across the borders. This small overhead is easily balanced by the efficiency improvements achieved by processing a large model in smaller sections. The most important gain results from a more compact memory access pattern.

In Monte Carlo radiative-transfer calculations the random walk nature of the path of a photon package makes it very difficult to attain a good cache hit rate if the size of the model is significantly larger than the CPU cache. Because of the significant latency caused by a cache miss, the computational speed is much lower than what is theoretically possible. The cache hit rate can be improved to some extent with special indexing schemes such as the space-filling Hilbert-Peano curve, but this has not yielded significant gains in speed owing to the additional computation required by the method \citep{Jonsson2006}. In comparison, using the hierarchial grid structure and grid-by-grid processing provides an efficient way to improve the cache hit rate and consequently also the computational speed. When the radiation is processed in one small grid at a time, a much larger fraction of memory accesses are to cached memory locations. Even in a model with only one hierarchy level (i.e., a single uniform grid), it can be advantageous to process the model in smaller parts. For instance, in a test calculation of light scattering in a molecular cloud, dividing a $512^3$ model into 512 $64^3$ cubes reduced the computation time to less than half. For an optimal performance, the grid size should be chosen so that the data for a single grid fits in the CPU L2 (or L3 if that exists) cache. For larger grid sizes, the cache hit rate drops quickly. On the other hand, a very large number of small grids should be avoided because of the increase in the overhead from boundary crossings. With typical cache sizes of a few megabytes, the optimal grid size is a few hundred thousand cells.

The programme needs to allocate several arrays for every grid in the model. For a grid with $N$ cells, the programme needs at least four separate arrays, each with $N$ elements: the density, the emission, the absorbed energy, and the subgrid identification.\footnote{A separate subgrid identification array is not strictly necessary as it is possible to encode the subgrid number in the unused parts of the density array.} If subiterations are used, one array for absorbed energy is needed for each level of hierarchy above the level of the current grid. Furthermore, for grids where ALI is used, all arrays for storing the absorbed energy need to be duplicated to store the absorptions from the same cell separately. In a model with a total of $N$ cells and $d$ levels and using 4 bytes for each array element, the worst case memory usage by the arrays is $4N(3+2d)$ bytes if the frequencies are simulated one-by-one. For a relatively large model with $10^7$ cells and 10 hierarchy levels this results in the worst case memory consumption of 920 MB. For real models, the memory usage is likely to be lower by a factor of at least five, because it is very unlikely that ALI is used in all grids and that a very large fraction of the cells are at the deepest hierarchy levels. Therefore, memory consumption by the grid arrays is usually not a serious limitation. In the case of polychromatic radiative-transfer, arrays for several frequencies need to be kept in the memory simultaneously, and for $f$ frequencies the worst case memory consumption is $4N[1+f(2+2d)]$. Memory limitations may restrict the number of frequencies that can be run simultaneously in large models. The grid-by-grid processing allows us to keep only the arrays belonging to the current grid in the computer memory. After completing a grid, the updated arrays can be written to the disc and the data for the next grid read into memory. As a result of the grid-by-grid processing, the number of times that the data for any grid need to be written or read is low. However, disc access should be kept to a minimum, and it is better to keep the whole model in the memory if possible.

In addition to the grid data, memory is needed for photon buffers that are used for communicating between different grids. The programme uses three 4 byte floats to store the position of the photon package, three floats for the direction of the package, and one float for the number of photons. If subgrid iterations are used, one 4 byte field is used to track the hierarchy level. If ALI is also used, one four byte field is needed to store the identity of the cell where the photon package started. A total of 36 bytes is needed for every photon package.\footnote{The memory usage could be easily reduced to 28 bytes per package, e.g., two floats would be enough to save the package direction and a single 4 byte field would be enough to save both the cell where the packet originated, and the hierarchy level.} In polychromatic Monte Carlo simulations each additional frequency requires 4 bytes. If a very low Monte Carlo noise level is necessary, the required number of photon packages can make the photon buffers too large to fit in the main memory. In that case, it is possible to store the photon buffers on the disc instead of the main memory. Another option is to send the photon packages to the model cloud in smaller batches. There is some overhead due to the need to start the photon transfer multiple times, but in practice the effect is negligible.

\section{Tests}
We have run tests to determine the accuracy of the programme, and the gains in computational cost that the new methods can provide. The dust model used in all tests is based on \citet{Draine2003}\footnote{See \url{http://www.astro.princeton.edu/~draine/dust/dustmix.html}.} and consists of a single dust component at the equilibrium temperature with the local radiation field. We use a relatively sparse frequency grid with only 36 frequencies stretching from the ultraviolet to the far-infrared. Frequencies are simulated one-by-one without using polychromatic packages.
\subsection{Molecular cloud}
In the first test, we compare the results of a full resolution simulation to results from a hierarchial model that has been generated by combining cells with a low optical thickness, forming grids from the combined cells and building a hierarchy tree from the grids. We examine by how much the accuracy reduces with the adaptive model, and the speed-up that is possible. In the case of a uniform grid programme, the code is identical to the programmme described in \citet{Juvela2003, Juvela2005}. That programme has been tested against several other radiative-transfer codes, and the results were found to be in excellent agreement (Baes et al., in preparation).

The model represents a part of a molecular cloud that is heated by the external radiation field and the stars embedded in the cloud. The density field is from a snapshot of an isothermal magnetohydrodynamic simulation with the Stagger code \citep{Padoan2007}. The simulation does not include self-gravity, and as a consequence the density contrasts in the model cloud are relatively low. We use as the full resolution model a $384^3$ piece from a $1000^3$ cube that was previously used in \citet{Lunttila2008, Lunttila2009}. The total number of cells in the model is thus $384^3\approx 5.7\times 10^7$.

The mean molecular hydrogen density is set to $n(\mathrm{H}_2)=70$ cm$^{-3}$ and the size of the grid to 2 pc. The mean $V$-band optical thickness through the cloud is only $\approx 0.4$, but along some lines of sight it reaches values of more than 50. The external radiation is described by the interstellar radiation field from \citet{Mathis1983}. The internal radiation sources are ten stars that are represented by blackbodies with temperatures between 7000 K and 9500 K and radii between 1.2 R$_\odot$ and 2.1 R$_\odot$. The stars are located in regions of high density.

The hierarchial model has a $48^3$ root grid and three levels of refinement for a maximum effective resolution of $384^3$. The model has 162 grids that contain a total of $4.1\times 10^6$ cells, which is lower by a factor of almost 14 than for the full resolution grid. The hierarchial model is built from the full grid by requiring that the $V$-band optical thickness of any combined cell cannot be larger than 0.05, and that the density contrast $\rho_{\mathrm{max}}/\rho_{\mathrm{min}}$ between the cells merged into one is always less than 30.

The model cloud has a relatively low maximum optical thickness and the dust temperature remains below 100 K. As a result, the cloud is optically thin to its own thermal dust emission and the dust temperature is almost fully converged after only one iteration. We therefore do not use subiterations in this model. We instead focus on the speed and accuracy of the formal solution algorithm in the full resolution and the hierarchial cloud models.

We compare the results using surface brightness maps of the cloud computed at the full $384^2$ resolution in directions parallel to the coordinate axes. The surface brightness calculated from the full-resolution model and from the hierarchial model are shown in Fig. 2, and the relative difference
\begin{equation}
\Delta=(I_{\mathrm{AMR}}-I_{\mathrm{full}})/I_{\mathrm{full}}
\end{equation}
between the results from the full model and from the hierarchial grid is shown in Fig. 3. The relative mean error
\begin{equation}
\Delta_{\mathrm{mean}}=\frac{\sum_{\mathrm{map}} (I_{\mathrm{AMR}}-I_{\mathrm{full}})} {\sum_{\mathrm{map}} I_{\mathrm{full}}}
\end{equation}
is small, less than $10^{-3}$ at all wavelengths. The relative root-mean-square (RMS) error
\begin{equation}
\Delta_{\mathrm{RMS}}=\sqrt{\frac{\sum_{\mathrm{map}} (I_{\mathrm{AMR}}-I_{\mathrm{full}})^2} {\sum_{\mathrm{map}} I^2_{\mathrm{full}}}}
\end{equation}
is larger, reaching values up to 0.04.

The error corresponds to positions with a steep density gradient, and where the hierarchial grid has a lower resolution than the full model. In the surface brightness maps that were calculated with the hierarchial model, the effect of the large cells can be seen as large areas of constant brightness (the large cells seen in projection). Their surface brightness is close to the mean brightness of the full resolution model over that region. However, because of the brightness gradient within the area, the surface brightness calculated with the hierarchial model is higher on one side and lower on the other than in the results from the full resolution model. These errors cancel each other almost exactly, and the mean error is very low. The effect can be easily seen by comparing Figs. 2 and 3. Bands of large relative error (both positive and negative) can be seen in areas with steep surface brightness gradients.

\begin{figure*}
  \includegraphics[width=17cm]{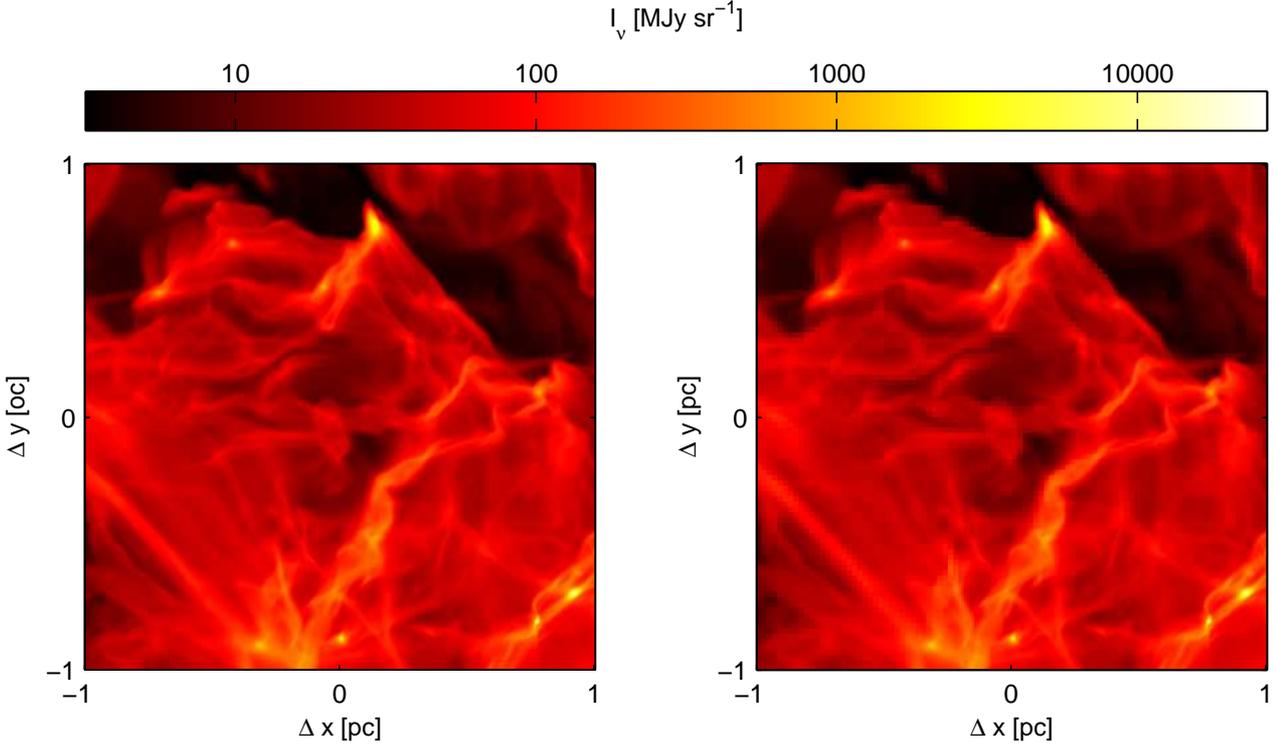}
  \caption{Surface brightness of the molecular cloud model at 100 $\mu$m. The left panel shows the map from the full resolution model and on the right we present the results of the adaptive model.}
\end{figure*}

\begin{figure*}
  \includegraphics[width=17cm]{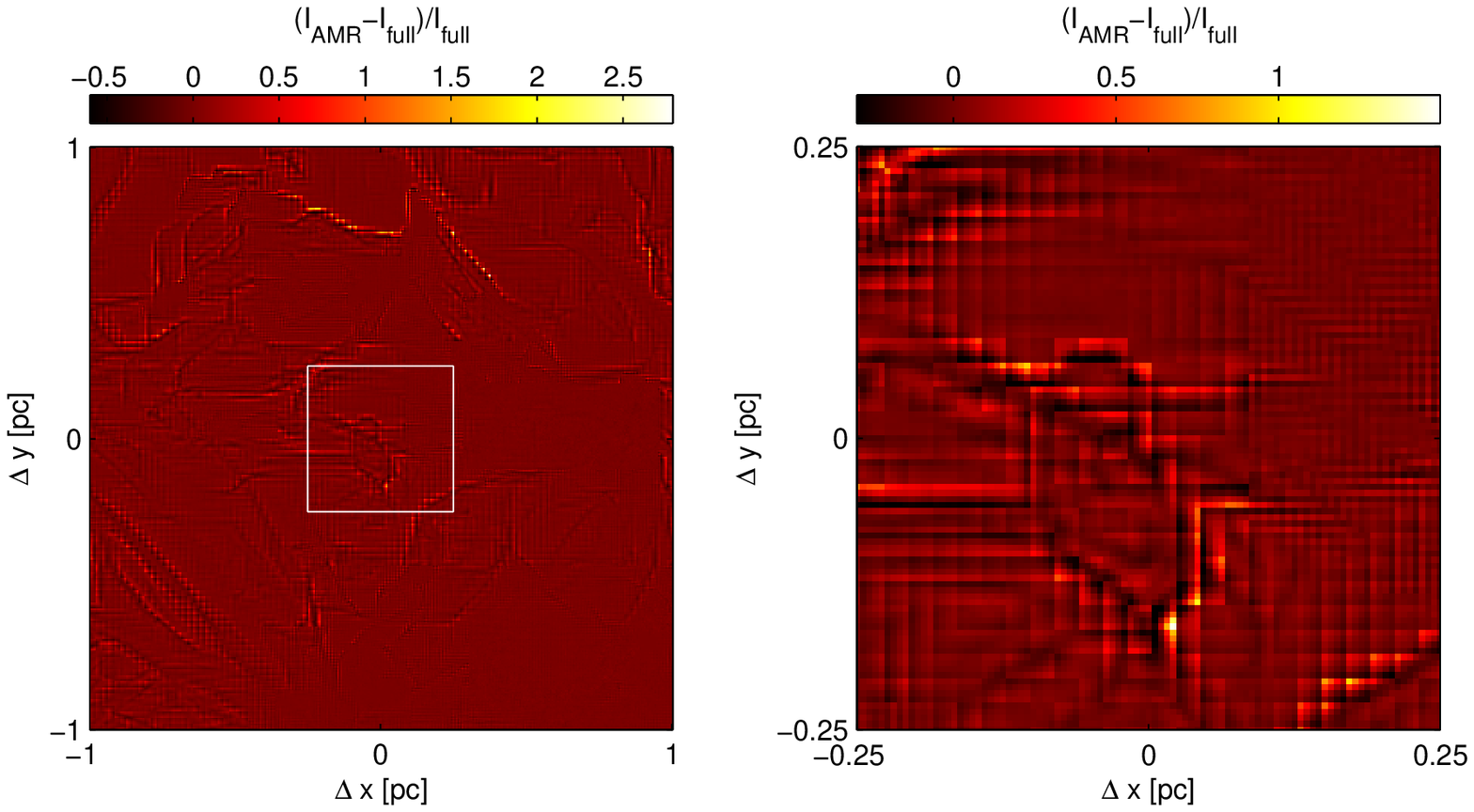}
  \caption{The left panels shows the relative difference of the molecular cloud's 100 $\mu$m surface brightness between the full-resolution model and the hierarchial model. The right panel shows an enlarged view of the region marked with the white box.}
\end{figure*}

The error depends crucially on how the hierarchial grid is constructed. If it is necessary to model steep intensity gradients accurately, the adaptive grid should be constructed with a small value of the largest allowed density contrast for merging the cells. On the other hand, for the calculation of the spectral energy distribution, this is not as important. In the present case, the high density regions were included in the adaptive grid with their full resolution because of the low allowed maximum cell optical thickness. As a result, the accuracy in the regions with a $V$-band optical thickness $>5$ is good with a relative RMS error $\approx 10^{-4}$.

When the same number of photon packages was used in both the hierarchial and the full resolution model, and no resampling (i.e., package splitting or Russian roulette) was done at the grid boundaries, the runtime of the hierarchial model was lower by a factor of 6.1. In the hierarchial model, resampling by a factor of four at the grid boundaries decreased the time spent in the computation of the radiation field produced by the stars by a factor of 1.8, while the calculation of the effect of the external radiation was slower by a factor of 2.3. However, resampling a smaller number of initial packages is sufficient to ensure an adequate sampling of the radiation field. Without resampling, a very large number of photon packages would need to be sent to get a good sampling of the radiation field at the deepest hierarchy levels, requiring a time consuming computation. The largest cells would then be crossed by a large number of packages, and the sampling noise in them would be much lower than in the smallest cells. If the number of photon packages sent to the model is chosen so that the desired noise level is reached in the small cells, the larger ones are sampled with an unnecessarily high precision, wasting computation time. With package splitting, a similar noise level can be reached at all levels.

To test how the resampling affects the noise level, we ran the Monte Carlo simulation of the 4 $\mu$m radiation field 20 times both with and without resampling, using the same relatively low number of initial photon packages. The results were compared to the reference solution, which was taken to be the mean of the solutions of all individual runs. We calculated the RMS difference between the reference solution and the results of individual runs at each level of the hierarchy
\begin{equation}
\delta(i)_{\mathrm{RMS}} = \sqrt{\frac{1}{N_{\mathrm{cells},i}}\sum_{k=0}^{N_{\mathrm{cells},i}} \left(J(k)-J(k)_{\mathrm{ref}}\right)^2},
\end{equation}
where $N_{\mathrm{cells},i}$ is the number of cells on level $i$ and the sum is taken over all cells in level $i$ grids. The calculations were done separately for the external radiation and the radiation from the embedded stars. Table 1 shows the mean difference between the reference solution and the individual results from 20 runs. In both cases, the noise level is normalised to the maximum value found at any level either with or without resampling.

The results of the test with external radiation show a significant improvement for the smallest grids where the noise level is highest. Without resampling, the noise level is higher by a factor of $\approx 6$, and to reach the same noise level without resampling, the number of packages would have to be increased by a factor of more than 30. Although the simulation of a single package sent to the model is slower when the resampling is used, the gain in computational speed is a factor of ten. In the case of the embedded stars, the gain is more modest. The noise levels in the most poorly sampled grids are similar with and without package splitting, and the gain is only due to the increased speed of simulating a single package. In the test case, the speed-up is less than a factor of two.

\begin{table*}
 \caption{Effect of package splitting on the noise level in the molecular cloud model.}
 \label{table:1}
 \centering
 \begin{tabular}{c c c c c }
 \hline\hline
  & \multicolumn{4}{c}{RMS noise level (arbitrary units)} \\
  & \multicolumn{2}{c}{External Radiation} & \multicolumn{2}{c}{Internal Sources} \\
 Hierarchy Level & With Resampling  & No Resampling   & With Resampling  & No Resampling \\
 \hline
     0 & 0.14   & 0.13  &    0.28  &  0.03 \\
     1 & 0.14   & 0.25  &    0.32  &  0.10 \\
     2 & 0.16   & 0.50  &    0.58  &  0.29 \\
     3 & 0.17   & 1.00  &    1.00  &  0.95 \\
 \hline
 \end{tabular}
\end{table*}

\subsection{Circumstellar disc}
To study the benefits of subiterations, we use a model of a flared circumstellar disc around a protostar. We adopt the same density structure as \citet{Pinte2009}
\begin{equation}
\rho(r,z)= \rho_0\left(r/r_0\right)^{-\alpha}\exp\left[-\frac{1}{2}\left(\frac{z}{h_0(r-r_0)^\beta}\right)^2\right],
\end{equation}
where $r_0=100$ AU, $h_0=10$ AU, $\alpha=2.625$, and $\beta=1.125$. The inner and outer radii of the disc are 0.1 AU and 400 AU. Parameter $\rho_0$ is chosen so that the midplane $I$-band optical thickness is $5\times 10^4$. The disc is illuminated by a protostar that is represented by a blackbody with a temperature $T=4000$ K and a radius of two solar radii. We note that although the model density structure and the radiation source are the same as in the benchmark tests of \citet{Pinte2009}, the dust model is different, and the results cannot be directly compared.

Although the density structure is in this case cylindrically symmetric, the disc is gridded into an adaptive 3D cartesian model with a $126\times 126\times 36$ root grid, corresponding to physical dimensions $800\times 800\times 240$ AU. The model has nine levels of refinement, resulting in the maximum effective resolution of $64000\times 64000\times 19200$, or a linear resolution of 0.0125 AU. The number of cells in the model is approximately $8.5\times 10^6$.

Because the inner edge of the circumstellar disc is only 0.1 AU from the protostar, it is heated to a temperature of several hundred kelvin. Consequently, much of its thermal emission is at the optical and near-infrared wavelengths. The optical thickness is very high in the inner parts of the disc at these wavelengths, and the dust thermal radiation cannot escape directly. Therefore, the $\Lambda$ iteration is expected to converge very slowly. Because the outer disc is cooler and has a lower density, it is optically thin to its own thermal emission and the iteration converges quickly. Thus, the model is well-suited to testing the subiteration algorithm.

We compare the accuracy and speed of the adaptive subiteration algorithm to those of the basic method of iterating the full model at every step. In this test, the automatic subiteration algorithm is used to select the grids that should be included in each iteration step. The criterion used for selecting the subsequent grids is the mass-weighted average of the difference between the emitted and absorbed energy, and all grids where the average is at least 0.25 times the maximum value found are chosen for the next iteration. The process is terminated after 30 steps, although the temperature distribution is not yet fully converged. We do not use ALI in this test.

Figure 4 shows the CPU time used as a function of the iteration step by both the full iteration and the subgrid algorithm. For the subgrid method, the whole model is not included in any iteration and in eight steps only the innermost grid is included. Figure 5 illustrates the convergence of the temperature in the disc mid-plane at different distances from the central star. As the outer grids are not included in every iteration, the temperature further from the star is updated only in the iterations marked with crosses. The algorithm starts by iterating only the innermost grids with the hottest dust, where the energy imbalance is initially the largest. Only in the later part of the calculation are the outer grids included. We note that the temperatures are those of individual cells, not azimuthal averages, and there is some Monte Carlo noise visible in the temperatures.

Figures 6 and 7 show the temperature distribution in the xy- and xz-midplanes (parallel and perpendicular to the disc, respectively) at the end of the iteration. Only a small part of the model, corresponding to the three deepest hierarchy levels near the central star, is shown. The images are calculated with the subiteration algorithm, but the results from the full iterations are very similar. The RMS temperature difference between the results at the end of the calculation was 2.0 K, while the mass-weighted RMS difference was 1.3 K. As the final RMS temperature was 76 K, the relative error was 0.026 (44 K and 0.028 with the mass-weighting). The discrepancy is largely due to Monte Carlo noise, because the difference between the final results of two runs with the subiteration algorithm was similar. Futhermore, the mean difference in the mean temperatures was only 0.03 K. In the largest grid that was not included in any subiteration, the final mean temperature after the full iterations is 0.14 K higher than in the subiteration results. If the subiteration algorithm is forced to run a single full iteration at the end, the bias disappears.

In this test the speed-up achieved by using subiterations is approximately a factor of two. The relatively modest gain is due to two things. Firstly, in the model a relatively large fraction of the cells are at the deepest levels of the hierarchy, the innermost grid having approximately 10 \% of the total cells. Secondly, the emission is strongly concentrated in the innermost grids. To improve sampling, the number of photon packages sent from a cell at each frequency is weighted by the total emission at that frequency from the cell so that each package starts with the same number of photons. The hot dust near the star emits much more energy than the cooler outer parts, and therefore most of the photon packages are sent from the few innermost grids. In the subiterations, the number of photon packages a cell sends is the same that it would send in a full iteration. Therefore, the total number of photon packages is almost the same in the subiteration of only the innermost grid as it is in the full iteration. The reduced time taken by the subiterations is in this case mostly due to the shorter time taken by the simulation of a package in a small subtree. Part of the acceleration is attributed to the reduced number of calls to the dust model evaluation, but with the simple dust model used in the test less than 5 \% of the total time is spent in that part of the simulation in the case of full iterations.

Figure 8 shows the image of the circumstellar disc in $V$-band as seen from 2 degrees above the plane of the disc. The central star is not seen directly because of the very high extinction in the disc, and the surface brightness is solely due to the scattered star light. Emission from the heated disc is at this wavelength less than $10^{-4}$ times the stellar emission. We examined the effectiveness of the accelerated peel-off algorithm by calculating images of the scattered light and comparing the running times with the conventional peel-off method. When images of the scattered light were computed towards five observers, the accelerated peel-off with precalculated extinction tables for the four smallest grids provided a speed-up by a factor of 3.8. In calculations with a larger number of observing directions, the speed-up can be even more significant, because the peel-off calculations take a larger fraction of the computing time. For the molecular cloud model, the accelerated peel-off yielded a speed-up of only 1.7 for five observing directions. The smaller gain is due to the small size of that model and the small number of hierarchy levels. In such a case calculating the extinction to the closest grid border is often almost as expensive as determining the extinction to the outer border of the model.

\begin{figure}
  \resizebox{\hsize}{!}{\includegraphics{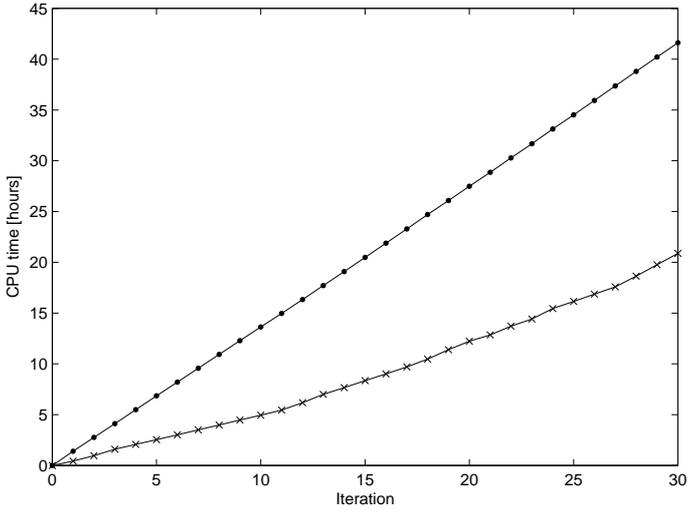}}
  \caption{Cumulative CPU time used by the simulation of the disc model as a function of iteration number for the full iteration (dots) and subiterations (crosses).}
\end{figure}

\begin{figure}
  \resizebox{\hsize}{!}{\includegraphics{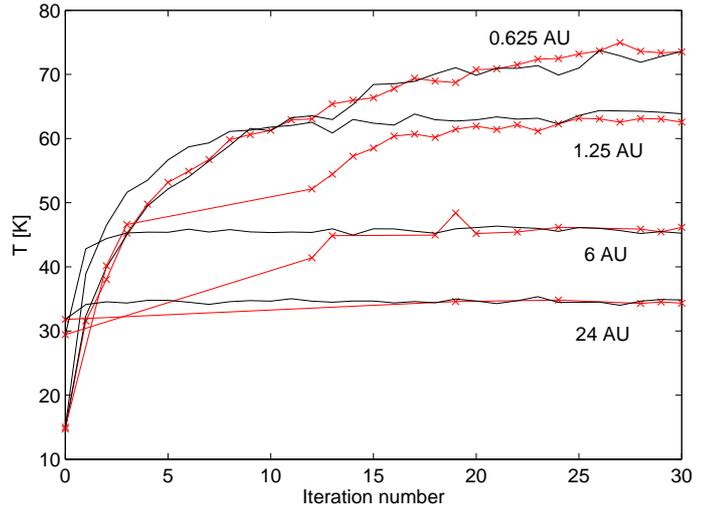}}
  \caption{Convergence of the temperature at cells near the xy-midplane of the disc model. The red line shows the results from the subiteration algorithm, and the crosses indicate the iterations where each grid was updated. The results of the full iterations, where all grids were updated at each iteration, are shown with the black lines without markers. The cells are approximately 0.625 AU, 1.25 AU, 6 AU, and 24 AU from the central star. }
\end{figure}

\begin{figure}
  \resizebox{\hsize}{!}{\includegraphics{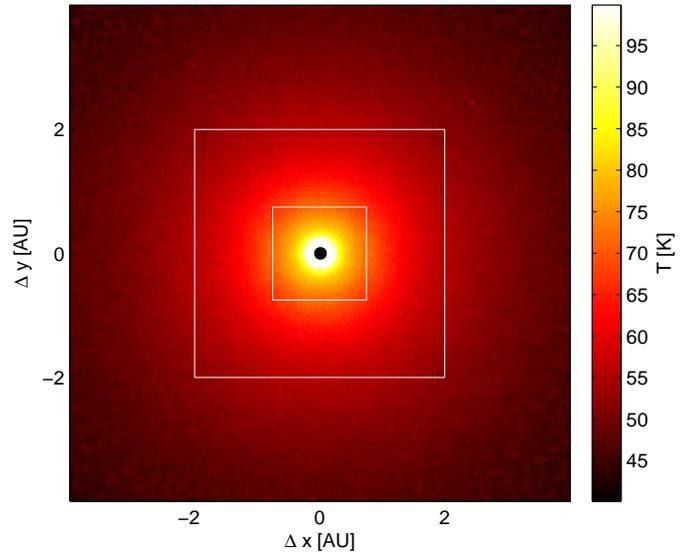}}
  \caption{Temperature distribution in the xy-midplane of the circumstellar disc model. The figure shows only an 8 AU$\times$8 AU part at the centre of the $800 \times 800\times 240$ AU model. The boxes indicate the boundaries of subgrids.}
\end{figure}

\begin{figure*}
  \includegraphics[width=17cm]{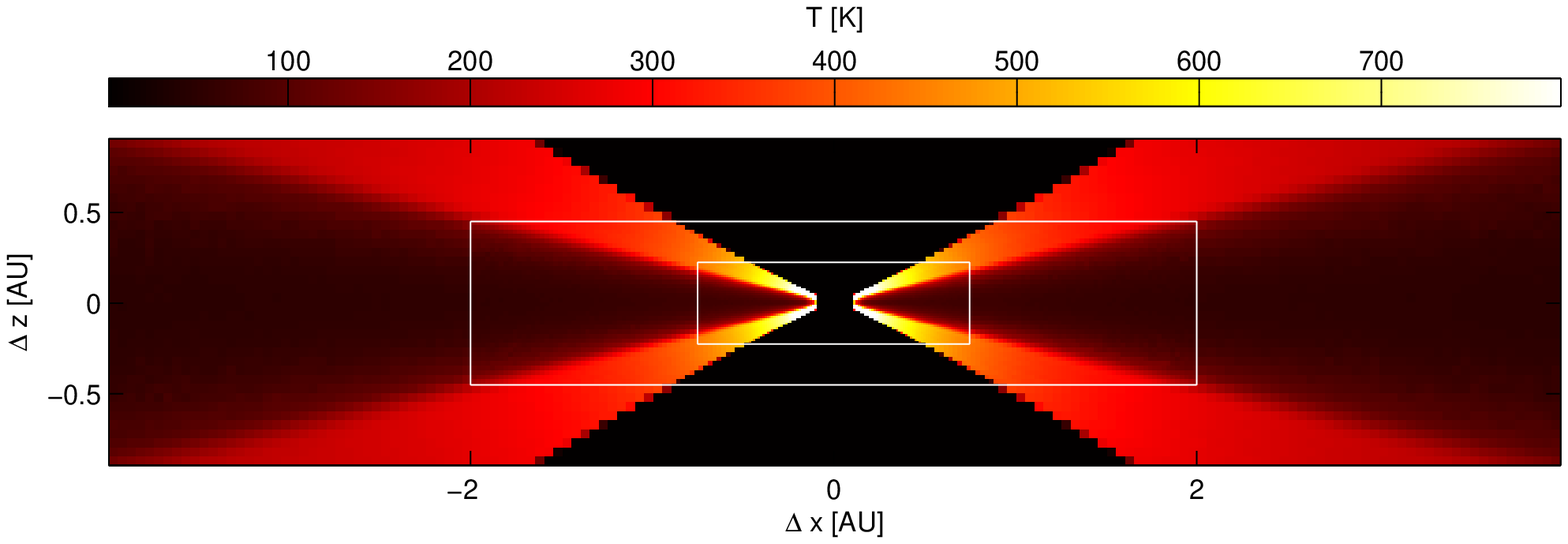}
  \caption{Temperature distribution in the xz-midplane of the circumstellar disc model. The figure shows an 8 AU$\times$1.8 AU part at the centre of the $800 \times 800 \times 240$ AU model. The boxes indicate the boundaries of subgrids.}
\end{figure*}

\begin{figure}
  \resizebox{\hsize}{!}{\includegraphics{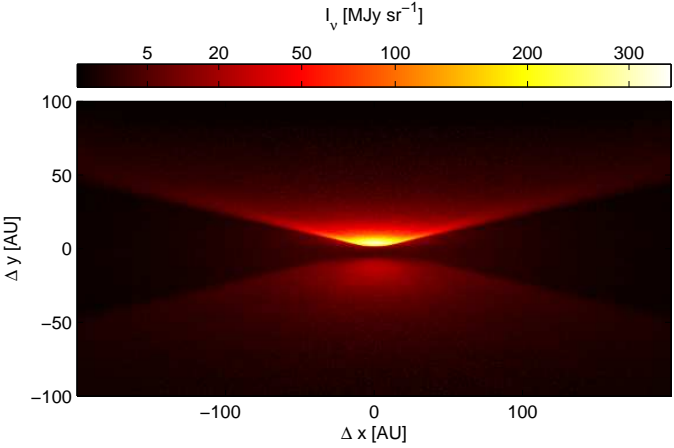}}
  \caption{Circumstellar disc seen in the $V$-band. The surface brightness is due to the scattered light from the central star that is hidden by the disc.}
\end{figure}

\section{Future work}
\subsection{Extending the method}
The use of subiterations does not depend on the method that is used to solve the linear radiative-transfer problem in the subtree. Although our programme uses the Monte Carlo method on cartesian 3D grids, in principle any coordinate system or method could be used. Some of the simplest changes use different coordinate systems in different grids. Near a small-sized radiation source, such as a star embedded in a molecular cloud, spherical coordinates are often better for describing the radiation field and the density structure, while the rest of the cloud can use cartesian grids. The computational cost of moving a photon package from a grid with the cartesian coordinate system to one that uses spherical coordinates or vice versa is not significantly larger than transforming between the cartesian coordinates of two different grids.

Instead of simple package splitting, a much more sophisticated resampling can be done on the boundaries. Because the entire radiation field crossing a grid boundary is known before the next grid is processed, it is possible to use a resampling that is adapted to the radiation field. For instance, if a very large number of packages are saved at the boundary, it may be enough to combine them and propagate a much smaller number of packages further. A more complicated change would be to use different solution methods for the linear problem in some of the grids. Because the grids only communicate through their boundaries, the method used to solve the problem can be different in each grid. The only requirement is that the solution method must be able to calculate the radiation flux leaving the grid from the given radiation field at the grid boundaries, and the emission from the sources inside the grid.

One possibility that could significantly accelerate the computation of optically thick grids is to explicitly calculate and store the entire $\Lambda$ operator for cells in the grid. As noted above, because the storage requirements increase as $O(N^2_{\mathrm{cells}})$, the method is practical only in small grids. However, it would allow the system in Eqs. \ref{rtequt} in the grid to be solved directly and very quickly with Newton-Raphson iteration. The solution should, of course, be recalculated possibly several times, if the radiation field entering the grid changes. For grids that are too large for this method, the currently existing ALI with a diagonal operator can be extended to more complicated approximate operators.

In some of the most detailed dust models, e.g., \citet{Compiegne2011}, the dust opacity is a function of temperature. This precludes using the radiation field from iterations of larger subtrees or from constant sources directly in subiterations. Even if the radiation sources stay constant, the radiation field can change because of the temperature-dependent opacity. Instead of separately storing the radiation from different hierarchy levels in each cell, the radiation must be saved in the \textit{inwards} array of each grid. As in the case of constant opacity, separate arrays are needed for storing the radiation field from different hierarchy levels. When a subtree is iterated, the radiation from the \textit{inwards} array of the subtree's root grid is transported into the subtree. In contrast to the previous case, radiation from the constant sources must be treated in the same way as emission from the medium. If the modifications allowing for the non-constant opacity were made, subiterations could also be used in line radiative transfer. The formal solution algorithm needs to be changed to take into account the line profiles and Doppler shifts, and the calculation of the dust emission has to be replaced with the solution of the rate equations. The use of subiterations is, however, identical to the case of dust radiative transfer with a temperature-dependant opacity.

\subsection{Parallelisation}
Monte Carlo radiative transfer can be easily parallelised within each iteration by dividing the photon packages between several processing cores. Each core can propagate the packages independently, and the results need to be combined only at the end to calculate the total radiation field in each cell. The programme is currently parallelised for shared memory architectures using OpenMP. The scaling is relatively good for a small number of cores. A speed-up by a factor of 4--6 is seen when using 8 cores, with the best gains being achieved for large models and when many photon packages are used. Similarly, calculating the dust emission for the next iteration using the dust model can be carried out in parallel. \citet{Robitaille2011} demonstrated good scaling to more than a thousand computational cores in Monte Carlo radiative transfer in distributed memory computers, especially if a large number of photon packages is needed.

While it is possible to parallelise each iteration as described above, the use of subiterations provides another route to parallel computation. Any subtrees that do not share grids can be processed independently. For instance, in the hierarchy shown in Fig. 1, subiterations with grids 1 and 2 as the root grids can be calculated  in parallel. In subtrees where a relatively small number of packages is sufficient, dividing them between different processes does not scale well to a large number of computational cores because the initialisation of the processes and the merging of the results takes a larger fraction of the total running time. In these cases, better scaling can be reached by processing several independent subiterations in parallel, and dividing each subiteration between a small number of processes. Only subiterations involving large subtrees cannot be run independently. However, as these subtrees require a large number of photon packages, dividing the photon packages between processes works well for these subiterations. The drawback of the method is that scheduling the iterations becomes more difficult.

If there is enough memory at each computational node for the whole model cloud and the required buffers, the methods described above can be used. As shown previously, this is likely to be the case in continuum radiative-transfer simulations. However, in line radiative-transfer computations several additional arrays, such as the components of the gas velocity field and the local turbulent linewidths, are necessary and it may not be possible to keep the whole model in each node's memory. In these cases, the model can be divided between different computational nodes according to the tree structure. Subtrees can be assigned to different nodes so that they can process radiation inside its subtree independently. It is only necessary to communicate the packages entering or leaving the subtree to different nodes. Because the algorithm already stores the packages at the grid boundaries, it is relatively easy to modify it to send the packages to another computational node in large batches instead of sending each package individually. Nevertheless, balancing the computational load between nodes is difficult. Furthermore, while in our tests with a small number of CPU cores disc I/O does not consume a significant fraction of the total running time, in a larger-scale parallel computation this may not be the case. We discuss the details of efficient parallelisation in a forthcoming article.

\section{Conclusions}
We have presented new algorithms for radiative transfer on hierarchial grids. We have tested the algorithms in realistic test cases and compared the results with existing methods. Our main conclusions are:
\begin{itemize}
\item The grid-by-grid processing provides some computational benefits owing to the higher cache hit rate. In Monte Carlo calculations, knowing the full radiation field at the grid boundary allows us to use adaptive resampling methods.
\item Results from a hierarchial model built from a uniform grid are close to the full resolution results, although the number of cells in the hierarchial model is smaller by more than an order of magnitude. Calculations with the hierarchial model are also faster than with the uniform grid.
\item Pre-calculated extinction tables can be used to accelerate the calculation of scattered flux. In a large model with a deep hierarchy structure, and when images of the scattered light are calculated for several observer directions, the speed-up can in practice be a factor of at least four.
\item Although in the circumstellar-disc test case using the subiterations only provided a speed-up of a factor of two, in other cases the gain can be far more significant. The subiteration algorithm is most beneficial in cases where the model contains several small, dense regions with a high optical depth embedded within a medium with a much lower mean density.
\end{itemize}
\begin{acknowledgements}
The authors acknowledge the support of the Academy of Finland grants 250741 and 132291. The authors thank the referee Dr. Doris Folini for comments that improved the paper.
\end{acknowledgements}

\bibliographystyle{aa}
\bibliography{hier}
\appendix
\section{Grid-by-grid processing}
As an example, we consider how the formal solution algorithm operates in the model depicted in Fig. 1. For the sake of simplicity, we assume that the Russian-roulette package termination and resampling at the grid boundaries are not used. The typical procedure is
\begin{enumerate}
\item The emission from internal sources in grid 5 is simulated with photon packages. The packages are followed until they reach the outer border of that grid, when they are saved in the grid's \textit{outwards} array.
\item The internal emission is simulated similarly in other grids. The packages that encounter a subgrid are saved in the subgrid's \textit{inwards} array. For example, a package saved from grid 3 may be saved in either the \textit{outwards} array of grid 3 or the 'inwards' array of grid 5. Grids at level 2 are simulated before grids at level 1, which are processed before the root grid. For grids on the same level, the order is arbitrary.
\item The packages that have reached the outer boundary of the root grid escape from the model volume and are removed from the simulation.
\item Packages are created in the root grid's \textit{inwards} array that describe the external radiation field impinging on the model volume.
\item The packages in the \textit{inwards} arrays are processed, starting from the root grid and moving inwards (i.e., level 0 before level 1, et cetera). The \textit{inwards} array of a grid is cleared after its packages have been simulated. We note that during the simulation of a grid, packages can only be entered into either the grid's \textit{outwards} array or the \textit{inwards} arrays of its child grids. Therefore, after the inwards packages for grid 5 have been simulated, the \textit{inwards} arrays of all grids are empty because of the top-to-bottom processing order.
\item The packages in the \textit{outwards} arrays are processed, starting from grid 5 and moving outwards. The packages in the root grid's \textit{outwards} array are removed from the simulation. At the end of this step, all the \textit{outwards} arrays are empty.
\item If there are packages left in any of the \textit{inwards} arrays, the algorithm returns to step 5. Otherwise, the computation is complete.
\end{enumerate}

\section{Subiterations}
\subsection{Implementation details}
When subiterations are used, we need to separately keep track of the radiation that enters the grid after visiting any of the higher levels of the grid hierarchy. Therefore, each grid has separate arrays for the radiation from the hierarchy levels above or equal to its own level. For example, a grid on level 2 has arrays for levels 0, 1, and 2. All photon packages keep track of the highest level of the hierarchy tree (closest to the root, i.e., the lowest level number) that they have visited. The package's contribution to the radiation field is stored in the appropriate array according to that level. For instance, consider a photon package that is created in a grid at level 2, e.g. grid 4 in the hierarchy shown in Fig. 1. If the package travels to grid 5, i.e. a grid at level 3, via a grid 2 that is at level 1, the package's contribution to the radiation field in grid 5 is added to the level 1 array of grid 5. The following subiterations start by calculating the dust emission using the latest radiation fields from previous subiterations. The total field in a cell is computed by summing the contributions from both the constant sources and the dust emission from different hierarchy levels
\begin{equation}
I_{\mathrm{tot}}= I_{\mathrm{stars}}+\sum_{i=0}^{i\leq \mathrm{level}} I(i),
\end{equation}
where $I(i)$ is the intensity stored in the level $i$ array.

Separate arrays for different hierarchy levels are needed to guarantee that the most recently calculated radiation fields are always used to compute the dust emission. As an example, we consider the subiterations in the hierarchy shown in Fig. 1. Suppose that after computing the radiation produced by the constant sources we run subiterations with subtrees under grids 0, 4, 2, 4, and 3 in that order. If the next subiteration is to be done with grid 0 again as the root grid (i.e., we iterate the whole model), we need to find the radiation field in every cell of the model to compute the dust emission. Determining the radiation field in grid 0 is simple: we can use the most recent field computed for that grid, i.e., the result of the first subiteration. Because the most recent subiterations did not include grid 0, they did not contribute to the radiation field there.

In grids deeper in the hierarchy, the situation is more complicated. If we simply used the most recently calculated radiation field for grid 3, i.e., the result from the fifth subiteration, we would ignore the effect of dust emission from all grids except 3 and 5. However, because we have earlier calculated subiterations with larger subtrees, we can use the results from these iterations for contributions from grids that were not included in the last subiteration. The level 1 array of grid 3 was last updated in the third iteration. It includes the effect of dust emission from grid 2, the parent of grid 3 and the effect of dust emission from grid 4 that had to enter grid 3 via grid 2. The array also includes the dust emission from grid 3 that exited the grid and finally scattered back. Similarly, the level 0 array was updated in the first iteration and contains the contributions from grids 0 and 1.

The subiterations are restricted to entire subtrees of the whole hierarchy to limit the amount of data that needs to be saved to track the contributions from different grids. If it were possible to iterate an arbitrary set of grids, it would be necessary to store for every cell in the model the radiation field that was emitted by each grid separately. For example, if in the model shown in Fig. 1, it were possible to run subiterations that include only grids 2 and 4, or grids 2 and 3, it would be necessary to store for each cell of grid 2 the effect of radiation emitted in grids 3 and 4 separately. In a small model such as the one shown in the figure, this could be done. However, a model can contain more than a thousand grids, and in larger models the extra memory consumption would be prohibitive. When the iterations are restricted to entire subtrees, the memory consumption scales with the number of hierarchy levels instead of the number of grids.

Restricting the subiterations to subtrees can result in a less efficient computation. If a grid needs to be iterated, all its descendants (i.e., children, children's children et cetera) are also included in the iteration. In practice, the grids at the deepest levels usually have the highest optical thickness and exhibit the slowest convergence, and it is rare that a grid at a high level in the hierarchy needs more iterations than its descendants. Another possible source of problems is that two neighbouring grids can communicate only in iterations that include a common ancestor grid (i.e., a grid whose set of descendants includes the neighbouring grids). Therefore, an optically thick region should be included in a single grid, or if it is divided between several grids, the common ancestor of the grids should not be far above them in the hierarchy. If the common ancestor is at a high level in the hierarchy, iterations of large subtrees are needed to communicate the effect of one part of the dense region to another. In some cases, avoiding this may not be possible without making the grids very large. However, we have not seen cases where this is a serious problem.

\subsection{A step-by-step example}
As an example of the subiteration method, we explain how the algorithm proceeds in the model shown in Fig. 1. We assume here that the subiterations are run with subtrees under grids 0, 4, and 2 in that order.
\begin{enumerate}
\item The radiation field due to constant sources such as stars is calculated and stored for each cell of every grid.
\item Thermal dust emission is calculated using the radiation field from the previous step as the input.
\item The radiative transfer simulation is run in the whole model (i.e., the subtree starting from grid 0) using the previously calculated thermal dust emission as the source. The calculated radiation field is stored in separate arrays according to the hierarchy level, as explained in the previous section.
\item Thermal dust emission is calculated again. This time the input is the sum of the radiation from the constant sources (from step 1), and the thermal emission from each level from step 3.
\item The radiative transfer is run in the subtree starting from grid 4, which in this case consists of only that grid. Only the radiation field stored for grid 4 can change, because no other grids are included. Furthermore, only the level 2 array of grid 4 can change. Because grids at higher levels of the hierarchy are not included, the photon packages cannot visit either levels 0 or 1 in this step.
\item Thermal emission from grid 4 is recalculated. In other grids, the radiation field has not changed, and it is not necessary to update the dust emission calculated in step 4. The input to the emission calculation is the sum of the radiation from the constant sources from step 1, the radiation in the level 0 and 1 arrays that was calculated in step 3, and the radiation in the level 2 array from step 5.
\item The radiative transfer is run in the subtree under grid 2. This step updates the level 1 array in grids 2, 3, 4, and 5, the level 2 array in grids 3, 4, and 5, and then the level 3 array in grid 5.
\item Dust emission from grids 2, 3, 4, and 5 is recalculated.

\end{enumerate}
\subsection{Efficiency of the algorithm}
The total CPU time taken by the subiteration algorithm can be written as
\begin{equation}
t_{\mathrm{sub}} = \sum_\mathrm{i=0}^{N_\mathrm{subiter}} \left[t_\mathrm{RT}(g_i)+t_\mathrm{dust}(g_i)\right],
\end{equation}
where $N_\mathrm{subiter}$ is the number of (sub)iterations and $t_{\mathrm{RT}}(g_i)$ and $t_{\mathrm{dust}}(g_i)$ refer to the time taken by radiative transfer and dust model evaluation step, respectively. The argument $g_i$ is the number of the root grid for the subtree processed in the $i$th iteration. For the special case where only iteration steps with the full model are taken, this yields
\begin{equation}
t_{\mathrm{full}} = \sum_{i=0}^{N_{\mathrm{full}}} \left[t_{\mathrm{RT}}(0)+t_{\mathrm{dust}}(0)\right] = N_{\mathrm{iter}} \left[t_{\mathrm{RT}}(0)+t_{\mathrm{dust}}(0)\right].
\end{equation}

The gain from using subiterations results from the smaller number of cells that need to be processed
\begin{equation}
\frac{\sum_\mathrm{i=0}^{N_\mathrm{subiter}}n_\mathrm{cells}(g_i)}{\sum_\mathrm{i=0}^{N_\mathrm{full}}n_\mathrm{cells}(0)} = R < 1,
\end{equation}
where $n_\mathrm{cells}(g_i)$ is the number of cells in the subtree under grid $g_i$. The reason that $R$ can be much less than one is that in full iterations the number of iteration steps needed is driven by the slowest converging regions, which often contain only a small fraction of the total cells in the model. With subiterations, these regions can be processed separately, and an optimal number of steps can be used for each.

Because the dust emission is calculated independently for each cell in the processed subtree, the time used in the dust model evaluation step is always proportional to $n_\mathrm{cells}(g_i)$
\begin{equation}
t_{\mathrm{dust}}(g_i)=C n_\mathrm{cells}(g_i),
\end{equation}
where $C$ is a constant that depends on the dust model. Therefore, the time spent in the dust model step with subiterations is reduced by a factor of $R^{-1}$ compared to the full iterations. If a complicated dust model is used, the computational cost of the radiative transfer step can be negligible compared to the cost of the dust model evaluation. In that case, the total speed-up from the sub-iterations is also $R^{-1}$.

For the radiative transfer step, the scaling with $n_\mathrm{cells}$ depends on both the details of the model and the algorithm used for calculating the formal solution. For example, raytracing with short characteristics on a uniform grid and with a fixed number of directions scales linearly with the number of cells, while for long characteristics the scaling is $n_{\mathrm{cells}}^{4/3}$.

For the Monte Carlo method, $t_{\mathrm{RT}}(g_i)$ is the product of the number of photon packages sent, $n_\mathrm{packages}(g_i)$, and the average time spent in the simulation of a single package, $t_{\mathrm{phot}}(g_i)$
\begin{equation}
t_{\mathrm{RT}}(g_i)=n_\mathrm{packages}(g_i) t_{\mathrm{phot}}(g_i).
\end{equation}
The number of photon packages depends both on the details of the model and the Monte Carlo sampling method employed. The number of packages is smaller for smaller subtrees, but not necessarily significantly. For instance, in the circumstellar-disc test case, most of the packets originate in the deepest hierarchy levels, and $n_\mathrm{packages}(g_i)$ is almost the same for all subtrees.

The average cost of tracking a single photon package is approximately proportional to the number of cells it needs to cross before exiting the cloud, or before the package is otherwise terminated. In the case of a uniform grid and non-scattering (or optically thin) medium, the number of cells encountered scales as $n_{\mathrm{cells}}^{1/3}$. In a hierarchial model with package splitting at the grid boundaries, Russian roulette, and a strongly scattering medium, the situation is more complicated, but the calculation is always faster in small subtrees.

In the circumstellar-disc test case, where the smallest grids contained a relatively large fraction of the total cells, $R\approx 0.26$ and the running time was reduced by a factor of approximately two. In other models, the gain from using subiterations can be much larger. For instance, \citet{Malinen2011} modelled a molecular cloud that had more than a hundred dense cores with embedded protostars. The grids containing the stars had together only 0.7 percent of the total cells in the model, but because of their high optical depths and hot dust they needed a large number of iterations. Computing one full iteration took more than 22 times longer than taking one iteration step in each of the grids that contained a star.
\end{document}